Gibaek Kim, and Jungho Kim*

# Efficient nanophotonic devices optimization using deep neural network trained with physics-based transfer learning (PBTL) methodology

**Abstract:** We propose a neural network (NN)-based surrogate modeling framework for photonic device optimization, especially in domains with imbalanced feature importance and high data generation costs. Our framework, which comprises physics-based transfer learning (PBTL)-enhanced surrogate modeling and scalarized multi-objective genetic algorithms (GAs), offers a generalizable solution for photonic design automation with minimal data resources. To validate the framework, we optimize mid-infrared quantum cascade laser (QCL) structures consisting of two regions: active and injection, which have different levels of feature importance. The optimization targets include five key QCL performance metrics: modal gain, emission wavelength, linewidth, and effective injection/extraction energies. To address the challenge of multiple local optima in the output latent space, we integrate a deep neural network total predictor (DNN-TP) with a GA, enabling scalable and nature-inspired optimization. By replacing computationally expensive numerical simulations with the DNN-TP surrogate model, the optimization achieves a speed-up of over 80,000 times, allowing large-scale exploration of the QCL design space. To improve model generalization with limited data, we introduce PBTL, which transfers knowledge from a DNN core predictor (DNN-CP) trained on active-region structures. This approach yields a 0.69% increase in prediction accuracy, equivalent to a 50% reduction in training data requirements, and leads to generate more feasible device structure with 60% improvement in evaluation metric during optimization. Evaluation of optimized structures shows an average L1 distance of ~43 from the training dataset, confirming the surrogate model's capacity to generate novel, physically valid QCL designs beyond the training distribution.

## 1 Introduction

The recent advancement of artificial intelligence (AI) has significantly accelerated the design and optimization processes across various photonic systems. Deep neural networks (DNNs) have shown remarkable capability in capturing the complex nonlinear relationships between input parameters and output responses. As a result, there is growing consensus that DNNs are well-suited for modeling and optimizing photonic architectures, enabling both forward prediction and inverse design of system behavior [1]-[3].

Quantum cascade lasers (QCLs) represent one of the most challenging classes of photonic devices to optimize due to their highly sensitive, quantum-scale structural dependencies. Typically comprising 20–30 alternating quantum wells and barriers with nanometer or sub-nanometer precision, the energy states and wavefunctions in QCLs are strongly influenced by the thickness of each layer. This, in turn, affects critical device characteristics such as emission wavelength, longitudinal optical (LO) phonon scattering, and interface roughness scattering [4]-[7]. Predicting this behavior generally requires solving the Schrödinger equation, which is computationally intensive.

Metaheuristic algorithms such as genetic algorithm (GA) and particle swarm optimization have been widely used to optimize QCL structures under specific performance constraints [8]-[11]. GAs emulates the natural evolutionary process to explore a diverse range of structural candidates, often yielding a variety of valid design solutions. However, due to the need to evaluate a large population of candidate structures over hundreds of generations, these methods become prohibitively expensive when paired with traditional physics-based simulators. Moreover, given the multi-modal nature of QCL performance metrics such as modal gain, emission wavelength, and energy level alignment, the objective function landscape often contains numerous local optima. Escaping these local optima requires increasing the population size, which further exacerbates the computational burden.

To address this issue, we propose to replace the conventional numerical simulator with a DNN-based surrogate model that significantly reduces the computational cost of optimization. By training the surrogate model on data generated from the

---
*Corresponding author: **Jungho Kim**, Department of Information Display, Kyung Hee University, 26 Kyungheedae-ro, Dongdaemun-gu, Seoul 02447, South Korea, E-Mail: junghokim@khu.ac.kr
**First author:** Department of Information Display, Kyung Hee University, 26 Kyungheedae-ro, Dongdaemun-gu, Seoul 02447, South Korea, E-Mail: kgb9805@khu.ac.kr

Schrödinger equation, we construct a virtual environment that approximates the quantum behavior of QCLs. Integrating this surrogate model into the GA framework allows us to efficiently explore the vast design space (with over $10^{15}$ possible configurations) while preserving the natural evolutionary dynamics of the optimization process. Owing to its simplicity and high parallelizability, the surrogate-based GA achieves a speedup of 80,000 times over conventional approaches.

However, training an accurate surrogate model generally requires a large amount of labeled databases that become increasingly challenging in high dimensional design spaces. To mitigate this, we propose a novel transfer learning approach of so-called physics-based transfer learning (PBTL) that leverages domain-specific knowledge about QCL operation. Unlike generic transfer learning (GTL) methods [12], [13], PBTL utilizes a pre-trained model that captures the essential physical behavior of QCLs in a reduced latent space. This knowledge is then transferred to the target model, improving learning efficiency without sacrificing physical interpretability.

In this work, we present efficient training methodologies for DNN surrogate model of photonic device, which is governed by Schrödinger equation. By applying domain acknowledge of the devices, we first developed a surrogate model of a DNN core predictor (DNN-CP) trained on active-region structures, which has high representation for core QCL functionality. Then, two different transfer approaches of PBTL and GTL methodologies are applied to obtain a high quality of a DNN total predictor (DNN-TP) trained on active and injector regions of the QCL. The PBTL ensemble surrogate model has higher accuracy and showed high optimization success ratio compared to GTL ensemble model. In addition, we demonstrate that our surrogate-assisted GA can discover novel QCL structures that satisfy multi-objective constraints much faster than conventional methods.

## 2 Method

Figure 1(a) illustrates the overall flowchart of the proposed method. The process begins with the generation of a training dataset by numerically solving the Schrödinger equation using numerical algorithms. The input features for the training dataset are the thickness values of 24 superlattice layers in a QCL structure. The output features include the peak emission wavelength (λ), maximum modal gain ($G_{43}$), homogeneous broadening linewidth (Γ), effective injection energy ($E_{inj}$), and effective extraction energy ($E_{ext}$).

The five output features are strongly correlated to specific superlattice region-particularly injection and active regions. Figure 1(b) shows the spatial regions and their associated energy levels and wave functions. By computing energy level differences between wave functions, we can define key modalities such as the emission wavelength, effective injection ($E_{inj}$-$E_4$) and extraction energy ($E_1$-$E_{ext}$). Additionally, the optical modal gain spectrum is calculated using the eigenvalues and eigenfunctions obtained from the Schrödinger equation. Figure 1(c) presents an example of the final modal gain spectrum.

Since the structure of active regions contains essential information regarding QCLs operation, we first create training dataset composed of various active region configurations while keeping the injection layers fixed. Using this dataset, we trained a model that captures the core operating principles of QCLs. Figure 2(a) shows a surrogate model of the DNN-CP, which has 8-layer active region superlattice input features and five different output features. Afterward, a second training dataset consisting of full QCL structures with both injection and active regions considered is generated. This dataset is used to train comprehensive surrogate models of the DNN-TP, which are designated in Figs. 2(b)-(d). Notably, the DNN-TP is trained with either direct learning (DL) or transfer learning (TL) methodologies for comparison. Figure 2(b) shows a schematic diagram of the DNN-TP with DL, where the weights of the DNN-TP are trained without TL. Figure 2(c) and (d) show the network structures of DNN-TPs implemented with two different transfer learning methodology of GTL and our proposed PBTL, respectively. Figure 2(c) and (d) show schematic diagrams of the DNN-TP two different transfer learning methodology.

Once trained, the DNN-TP model is used as a surrogate simulator to evaluate candidate QCL structures during the optimization process. This dramatically reduces computational time while maintaining physical accuracy, enabling scalable and efficient exploration of high-dimensional design spaces.

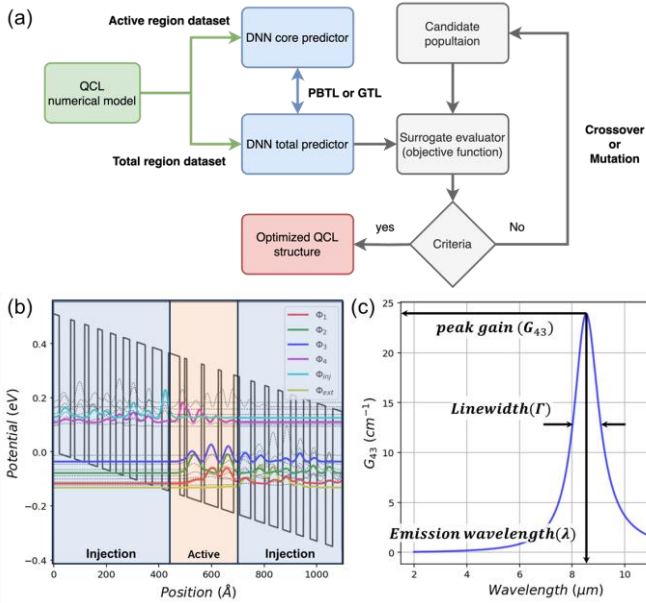

Figure 1 (a) Overall procedure of methodology, categorized to training dataset generation process (green), training of the DNN surrogate models (blue), optimization of the QCLs (gray), and optimization result (red). (b) The wave functions, energy levels and (c) modal gain spectrum of the given QCL superlattice **43**/18/**9**/55/**11**/53/**12**/47/**22**/43/**15**/38/**16**/34/**18**/30/**21**/28/**25**/27/**32**/27/**36**/25 with thicknesses are in angstroms. The InAlAs barriers' thicknesses are bold, while the InGaAs wells are in normal face.

## 2.1 Numerical modeling of QCLs

To generate high quality training dataset, various numerical algorithms were employed to accurately capture the physical behavior of mid-infrared (mid-IR) QCLs. In this study, we consider mid-IR QCLs structures, which are composed of alternating layers of $In_{0.53}Ga_{0.47}As/In_{0.52}Al_{0.48}As$ lattice-matched to an InP substrate. The conduction band (CB) discontinuity between the well and barrier materials is set to 0.51 eV. The total superlattice consists of 24 layers, with a four-well (4QW) active region and an eight-well (8QW) injection region, designed based on the double phonon resonance principle.

Given material properties, the initial potential is defined and used as an input to solve the Schrödinger equation. Because the upper lasing states in the active region are located far from the CB edge, the nonparabolicity (NPB) of the CB must be considered for accurate modeling. To this end, we employ an effective two-band model to incorporate CB nonparabolicity and solve the Schrödinger equation at the zone center [14]. To solve the equation, numerical QR algorithm is used.

From the calculated energy levels and wave functions, key physical output parameters such as a peak emission wavelength ($\lambda$), an effective injection energy level ($E_{inj}$), and an effective extraction energy ($E_{ext}$) are extracted. To compute the optical gain spectrum, the intersubband optical dipole moment, intersubband longitudinal optical (LO) phonon scattering rate [15], and intrasubband interface roughness (IFR) scattering rate [16]-[18] are determined using Simpson's 1/3 rule. The detailed theoretical modeling process for the mid-IR QCL is presented in Algorithm 1. Details of numerical modeling process is presented in Supplementary information (SI) section 1.

Figure 1(b) depicts the reference superlattice structure along with its corresponding energy levels and wave functions [14]. The optical transition between state 4 and state 3 is responsible for lasing while the three lower energy states ($E_1$, $E_2$, $E_3$), which are approximately separated by one LO phonon energy, facilitate efficient carrier extraction. The left-hand side of the injector region enables electron injection from the $E_{inj}$ into $E_4$ in the active region, and extraction from $E_1$ to $E_{ext}$ in the subsequent period. Owing to low effective injection and extraction energies, an intense population inversion is expected between states 4 and 3, resulting in high optical modal gain. As shown in Figure 1(c), the reference structure indeed exhibits a strong gain spectrum, confirming its optimized lasing behavior.

```
Algorithm 1 Mid-IR QCL Simulation Pipeline for Training Dataset Generation
 1: Input: Material parameters of $In_{0.53}Ga_{0.47}As/In_{0.52}Al_{0.48}As$
 2: Set: Conduction band offset $\Delta E_c = 0.51$ eV
 3: Define superlattice structure with 24 layers:
      • Active region: 4 quantum wells (4QW)
      • Transport region: 8 quantum wells (8QW)
 4: Generate initial conduction band potential profile $V(z)$
 5: Incorporate conduction band nonparabolicity using two-band effective mass model
 6: Solve: Schrödinger equation with nonparabolic conduction band
```
$$\left[-\frac{d}{dz}\left(\frac{\hbar^2}{2m^*(z)}\frac{d}{dz}\right) + V(z)\right]\psi_n(z) = E_n\psi_n(z)$$
```
 7: Use numerical QR algorithm to compute eigenvalues $E_n$ and eigenfunctions $\psi_n(z)$
 8: Calculate: Physical parameters
      • Peak emission wavelength $\lambda$
      • Injection level $E_{inj}$
      • Extraction level $E_{ext}$
 9: Evaluate: Optical gain spectrum
      • Compute intersubband dipole moment: $z_{nm} = \int \psi_n(z) z \psi_m(z)\, dz$
      • Compute intersubband LO-phonon scattering rate (Ref. [14])
      • Compute intrasubband interface roughness scattering rate (Refs. [15]–[17])
      • Use Simpson's 1/3 rule for numerical integration
10: Output: Energy levels, wavefunctions, optical gain, and related QCL parameters
```

Algorithm 1 Numerical modeling process of mid-IR QCL

**2.2 Training of the DNN surrogate models with PBTL**

The objective of training DNN model is to tune the weight of the NNs to predict output features of the given QCL structures. Figures 2(a) and (b) illustrate the NN structure of DNN-CP and DNN-TP. The size of training dataset plays a crucial role in determining the representation capability of the model. In general, as the dimensionality of the input feature space increases, significantly larger dataset is required to achieve a comparable level of accuracy. Given that generating training dataset involves computationally intensive quantum mechanical simulation, it is essential to maximize model generalization ability while minimizing training dataset size.

To this end, we first train the DNN-CP with DL method using a dataset focused on the active region, where physical phenomena most critical to lasing occur. DL means updating the network parameter using training dataset exclusively. In this dataset, the thicknesses of the injection layers are fixed while only the active region layers are varied. This dimensionality reduction effectively decreases the number of possible input combinations from $10^{15}$ to approximately $5 \times 10^9$. As a result, a compact dataset of 35,000 samples is sufficient to encode the fundamental quantum operation of QCLs into the DNN-CP. For training the DNN-TP, which predicts the behavior of the full QCL structure, we generate a dataset of 100,000 samples. This size reflects a practical trade-off between improved representational accuracy and the computational cost of dataset preparation (see SI section 2).

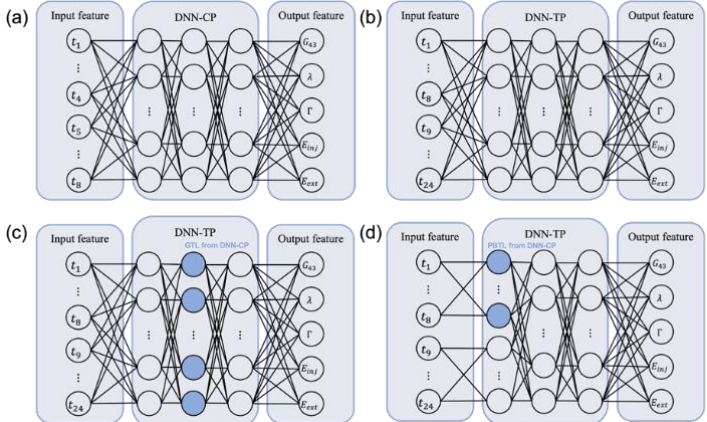

Figure 2 (a) Structure of the DNN-CP with 8-layer active region superlattice input features and five different output features, (b) structure of the DNN-TP based on DL with 24-layer total QCL structure and five different output feature, (c) structure of the DNN-TP based the GTL where the pretrained weights of the DNN-CP are transferred, (d) structure of the DNN-TP based our proposed PBTL of which the first hidden layer is divided into two parts that can process thickness of active and injector layers, respectively.

During dataset generation, input structures are initially sampled uniformly from predefined thickness ranges (see Table S1). However, since functioning QCL structures constitute a sparse subset of the entire design space, random sampling often results in non-operational devices. To address this, we employ a supplementary strategy: previously validated QCL designs are selected and randomly perturbed to generate additional realistic training examples that maintain population inversion conditions essential for laser operation.

To improve the representation ability of the model efficiently, we applied TL approach. Figure 2(d) illustrates the scheme of our proposed PBTL approach. We split the first hidden layer of the DNN-TP into two parts—corresponding to the active and injector regions—and transfer the pre-trained weights of the DNN-CP only to the active region block. For comparison, we also implement a GTL approach in Fig. 2(c), in which the second hidden layer of DNN-CP is transferred to DNN-TP without considering physical correspondence. In the case of the GTL, the weights of the first hidden layer of DNN-CP are not transferable due to the mis-match of the input data dimension (8 for DNN-CP and 24 for DNN-TP). Transfer learning and training details including NN structure and hyperparameters are presented in SI section 3 and 4.

### 2.3 Optimization of QCLs with surrogate-assisted GA

The optimization process begins by randomly generating QCL candidate structures within a predefined design space. Each candidate is evaluated using the DNN-TP surrogate model to predict key performance metrics. Based on the predicted objective values, high-performing candidates—referred to as elites—are selected as parent structures for crossover, producing offspring structures for the next generation. This process represents exploitation, as it leverages knowledge from previously successful designs. In parallel, exploration is maintained through mutation, in which new structures are randomly generated with a mutation rate of 10%.

Algorithm 2 illustrates the overall flowchart of this surrogate-assisted GA. The fitness of each candidate structure is evaluated using a composite objective function, which is defined as

$$f_1(G_{43}[cm^{-1}]) = \frac{1}{1+\exp\left(-0.5(G_{43}-G_{target})\right)} \quad (1)$$

$$f_2(\Gamma[meV]) = \frac{1}{1+\exp\left(0.5(\Gamma-\Gamma_{target})\right)} \quad (2)$$

$$f_3(\lambda[\mu m]) = \frac{1}{0.4\sqrt{2\pi}} \exp\left(\frac{-0.5(\lambda-\lambda_{target})}{(0.1)^2}\right) \quad (3)$$

$$f_4(E_{inj}[meV]) = \frac{1}{1+\exp\left(0.5(E_{inj}-E_{inj,target})\right)} \quad (4)$$

$$f_5(E_{ext}[meV]) = \frac{1}{1+\exp\left(0.5(E_{ext}-E_{ext,target})\right)} \quad (5)$$

$$f_{total}(t_i) = f_1 + f_2 + f_3 + f_4 + f_5 \quad (6)$$

In (6), $t_i$ denotes the vector of superlattice layer thicknesses. Each term in the objective function corresponds to a specific performance metric—modal gain, linewidth, emission wavelength, and injection/extraction energies—and is normalized to have a maximum value of 1, ensuring balanced evaluation across all objectives. Due to the multi-objective nature of the problem, the design space contains multiple local optima. To avoid premature convergence and ensure sufficient exploration, we generate 2,000 candidate structures per generation with a mutation rate of 10%. The algorithm is run for 50 generations, yielding a total of 100,000 evaluations per optimization run. Such a computational scale would be impractical if using conventional numerical solvers, as each evaluation would require solving the Schrödinger equation.

To address this challenge, we replace time-consuming numerical evaluation with the DNN-TP surrogate model, which provides fast and accurate predictions of QCL performance. This approach enables efficient and scalable optimization of complex QCL structures while preserving the benefits of nature-inspired evolutionary search.

---

**Algorithm 2** DNN-TP Assisted Genetic Algorithm for Mid-IR QCL Design Optimization

1: **Input:** Design space of layer thicknesses for $In_{0.53}Ga_{0.47}As/In_{0.52}Al_{0.48}As$ QCL structures
2: **Load:** Pretrained DNN-TP surrogate model for predicting QCL performance metrics
3: **Initialize:** Genetic Algorithm parameters
   - Population size $P$
   - Crossover rate $r_c$, mutation rate $r_m$
   - Number of generations $G$
4: **Generate:** Initial population of QCL structures $\{X_i\}_{i=1}^{P}$
5: **for** $g = 1$ to $G$ **do**
6:    **for** each individual $X_i$ in population **do**
7:       Predict QCL performance metrics using DNN-TP:

$$Objective(X_i) \leftarrow DNN\text{-}TP(X_i)$$

8:    **end for**
9:    **Select:** Parents via elite selection
10:   **Apply:** Crossover and mutation to produce offspring
11:   **Form:** New population using elitism and offspring replacement
12: **end for**
13: **Output:** Optimized QCL design candidates with high predicted performance

Algorithm 2 mid-IR QCL optimization with surrogate-assisted GA

## 3. Results and Discussion

**3.1 Impact of PBTL on surrogate model training**

Training a regression DNN model to accurately interpret complex physical phenomena is heavily influenced by the dimensionality of the input feature space. A higher input dimensionality implies a more complex mapping between input and output features, which in turn requires a significantly larger training dataset to achieve comparable accuracy. In our case, the DNN-CP and DNN-TP models differ substantially in their input dimensionality—8 versus 24, respectively. This increase leads to an estimated $10^{17}$ fold increase in the size of the design space, making the training of DNN-TP notably more challenging.

Figure 3(a) illustrates this trend: the minimum validation loss of DNN-CP is approximately an order of magnitude lower than that of DNN-TP (see supplementary for details). It is important to note that these models were trained on different dataset sizes—35,000 samples for DNN-CP and 100,000 samples for DNN-TP—reflecting the differences in problem complexity. Although increasing the dataset size improves the performance of DNN-TP, it is computationally impractical to generate significantly larger datasets due to the high cost of numerical simulations (see Supplementary Figure S1). To mitigate this challenge, we apply a TL approach, reusing the rich physical representations learned by the DNN-CP. Figure 3(b) presents the UMAP (uniform manifold approximation and projection) distribution of the input feature spaces for both datasets. UMAP is a dimensionality reduction technique that preserves the local and global structure of high-dimensional data, allowing intuitive comparison of latent distributions. As shown in Fig. 3(b), the input features of DNN-CP dataset are narrowly distributed in latent space. This indicates that the DNN-CP more effectively capture fundamental operation principle. In addition, since both models are trained under the same governing Schrödinger equation and share a similar input latent space, we hypothesize that transfer learning can effectively enhance the generalization capability of the DNN-TP.

Figure 4(a) compares the training and validation loss curves of DNN-TP trained using three methods: DL, GTL, and PBTL. The figure shows the three models can be well trained without severe overfitting problem. The minimum training loss follows the trend of PBTL < GTL < direct, indicating that both TL approaches help in learning the input–output mapping more efficiently than training from scratch. However, in Figure 4(b), the validation loss shows a different trend: PBTL < direct < GTL, suggesting that GTL, despite the improved training performance, negatively affects generalization. This degradation is further evidenced in the weight distribution analysis. In Figure 4(c), the weight distribution of the GTL-trained model deviates significantly from a normal distribution, implying that the transferred weights are misaligned with the target task and hinder the model's representational capability. In contrast, the PBTL-trained model maintains a stable and Gaussian-like weight distribution, supporting its superior generalization performance.

To quantitatively assess model performance, we calculate the Pearson correlation coefficient (PCC) and the coefficient of determination ($R^2$ score) for each output feature. Table 1 presents the evaluation metrics for models trained using three methodologies: DL, GTL, and PBTL. All models achieve a PCC greater than 0.90 across the five modalities, indicating reliable prediction performance even on unseen structures. Importantly, the PBTL-based model consistently outperforms the DL model across all output features, with the most significant improvement observed in modal gain prediction.

This result highlights the effectiveness of knowledge transfers from the DNN-CP, which is trained on a compact latent space where the active region design and modal gain are strongly correlated.

In the case of GTL, improvements in evaluation metrics are not significant, and some output modalities even show degraded performance compared to the DL baseline. This suggests that indiscriminate transfer of learned weights, without considering the underlying physical correspondence, may harm model generalization. In contrast, PBTL effectively transfers domain knowledge from DNN-CP, resulting in a 0.69% improvement in overall accuracy. Based on our learning curve analysis, this accuracy level would require training with approximately 150,000 samples if DL is used. Therefore, the proposed PBTL framework can reduce the computational burden of training dataset generation by nearly 50%, without compromising model performance (see Supplementary Figure S1).

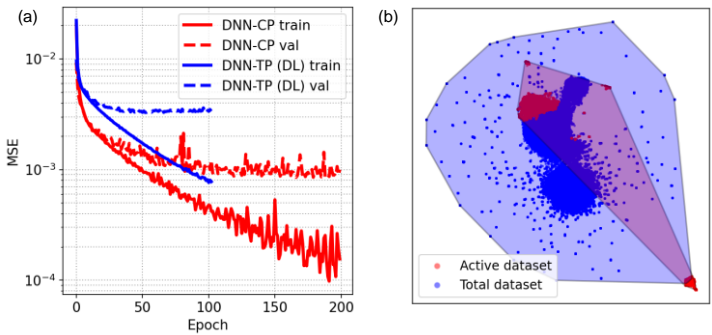

Figure 3 (a) Loss curve of two models to compare the performance between DNN-CP vs DNN-TP with DL. The solid lines represent the training losses and the dashed lines indicate validation losses. (b) Training dataset distrubtion of active dataset and total dataset, which are projected by UMAP.

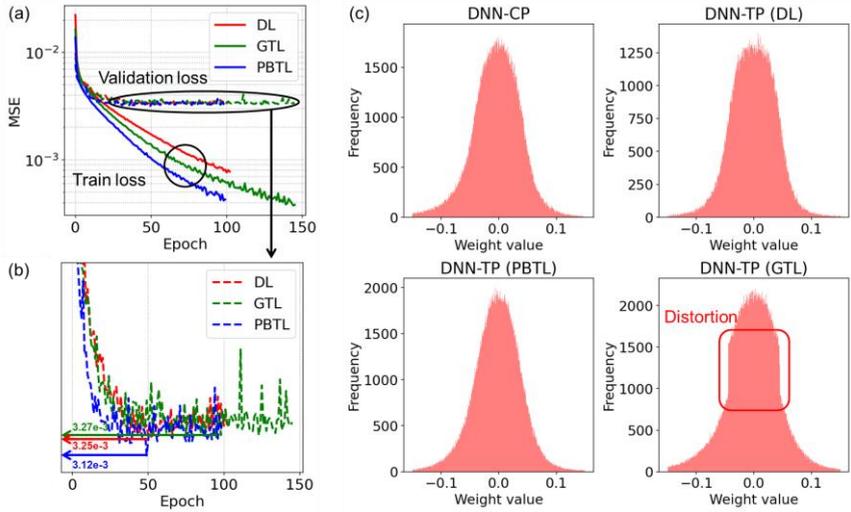

Figure 4 (a) Loss curves of DNN-TP models trained using three different methodologies of DL, GTL, and PBTL. Solid lines represent training loss while dashed lines indicate validation loss. (b) Validation loss trajectories and the minimum loss values for each model. (c) Distribution of weight values across the different models.

Table 1 PCC and R2 score of the output features for DNN-TP trained by three different methodologies.

| Output parameter | Direct learning | | Generic Transfer Learning | | Physics-Based Transfer Learning | |
| --- | --- | --- | --- | --- | --- | --- |
| | PCC | R2 | PCC | R2 | PCC | R2 |
| Modal Gain | 0.913 | 0.831 | 0.916 (+0.39%) | 0.837 (+0.73%) | 0.922 (+0.99%) | 0.850 (+2.19%) |
| Linewidth | 0.890 | 0.789 | 0.886 (-0.48%) | 0.780 (-0.03%) | 0.891 (+0.11%) | 0.792 (+0.47%) |
| Emission wavelength | 0.966 | 0.932 | 0.967 (+0.15%) | 0.935 (+0.28%) | 0.967 (+0.17%) | 0.935 (+0.36%) |
| Effective injection energy | 0.984 | 0.968 | 0.984 (-0.03%) | 0.967 (-0.06%) | 0.985 (+0.08%) | 0.969 (+0.15%) |
| Effective extraction energy | 0.990 | 0.979 | 0.992 (+0.19%) | 0.984 (+0.54%) | 0.992 (+0.17%) | 0.984 (+0.47%) |
| Average | 0.948 | 0.900 | 0.949 (+0.04%) | 0.901 (+0.09%) | 0.951 (+0.29%) | 0.906 (+0.69%) |

### 3.2 Utility of DNN surrogate model

In this section, we evaluate the DNN-TP surrogate model, which is trained with PBTL methodology, in terms of both prediction accuracy and computational efficiency, as required for its integration within the optimization framework. To be useful during multi-objective optimization, the surrogate model must reliably predict key QCL performance metrics across a diverse range of input structures.

To assess the model's generalization ability, the DNN-TP trained with the proposed PBTL methodology is tested on a separate test dataset, unseen during training. Figure 5(a) shows scatter plots comparing the DNN-TP predictions with ground-truth values obtained from numerical simulations. A strong linear correlation is observed across all output features, confirming the model's ability to accurately infer key QCL behaviors.

However, some deviations from the ideal linear trend are noticeable, particularly for more intricate output features such as modal gain and linewidth. This can be explained by the physical complexity of these properties. While the emission wavelength and effective energy levels (injection and extraction) are directly determined by the energy level differences in the superlattice, the linewidth and gain are affected by additional mechanisms—such as IFR and LO-phonon scattering—which involve interactions among multiple energy states and wavefunctions. This added complexity results in relatively higher prediction errors for these two modalities.

During the QCL optimization process, the surrogate model is used to evaluate thousands of candidate structures in each generation. Given the multi-objective nature of the task, the solution space contains many locally optimal structures that satisfy only a subset of the design targets. To escape these local optima and ensure sufficient exploration, it is necessary to evaluate a large population of candidates. However, applying conventional numerical solvers to such a large design pool is infeasible due to their serial nature and high computational cost. The DNN-TP surrogate model addresses this limitation by leveraging matrix multiplication and vectorized operations, which are highly parallelizable. As shown in Figure 5(b), the computational time of the numerical solver increases linearly with the number of evaluated structures, requiring approximately 160 seconds to evaluate 100 QCL candidates. In contrast, DNN-TP evaluates the same number of structures in only a few hundred milliseconds, and its runtime remains effectively constant.

This efficiency gain translates into a dramatic acceleration of the optimization process. For example, in a GA configured with 50 generations and 2,000 candidates per generation, a numerical solver would require approximately 160,000 seconds (~44 hours) to complete a single optimization run. In contrast, the surrogate-assisted GA completes the same process in just a few seconds. Overall, the use of the DNN-TP surrogate model leads to a speedup of ~80,000 times while maintaining comparable accuracy to full-physics simulation. This allows for extensive design space exploration and effectively mimics the natural evolutionary process underlying genetic algorithms.

### 3.3 QCL Optimization with surrogate-assisted GA

#### 3.3.1 Search Space Convergence and the Role of Surrogate Modeling

The optimization is carried out with the following target objectives: a modal gain of 25 cm$^{-1}$, a spectral linewidth of 24 meV, and injection ($E_{inj}$) and extraction ($E_{ext}$) energy level differences of 26 meV each. Figure 6(a) illustrates the distribution of QCL candidate structures across the optimization process. Initially, the population is widely dispersed across the latent design space. However, as optimization proceeds, the population gradually converges into a narrow region around an optimal solution. This implies that to find the optimal structure, the algorithm must be capable of identifying such a narrowly confined region.

Given that GA is a stochastic optimization algorithm, it carries the inherent risk of converging to local optima if exploration capability is insufficient. Since we observe the optimal structures lies within extremely narrow region in latent space, an extensive population must be evaluated to ensure proper convergence in multi-objective optimization. Figure 6(b) presents the impact of population size on optimization performance. The aggregated objective score reaches its maximum value (~5) when the population size is increased to 2,000 per generation. While the increment of population promotes better global optimization ability, it also requires significantly more computational resources, particularly if numerical solvers are used for evaluation. This challenge motivates the use of a DNN surrogate model to enable more efficient exploration.

#### 3.3.2 Impact of PBTL approach on Optimization Outcomes

Since surrogate predictions do not guarantee physical validity, we validate the optimized structures using a physics-based numerical simulator. To quantify model effectiveness, we compare the pass@128 ratio—defined as the number of physically valid structures among the top 128 candidates—between DNN-TP models trained via DL and PBTL. Figure 6(c) reveals that even a small difference in model accuracy can result in large disparities in optimization outcomes. Notably, although the DNN-TP-assisted GA shows a high pass ratio initially, the ratio declines as the population size increases. This can be attributed to the reduced selection pressure in large populations, allowing more diverse but potentially unreliable candidates to pass through generations.

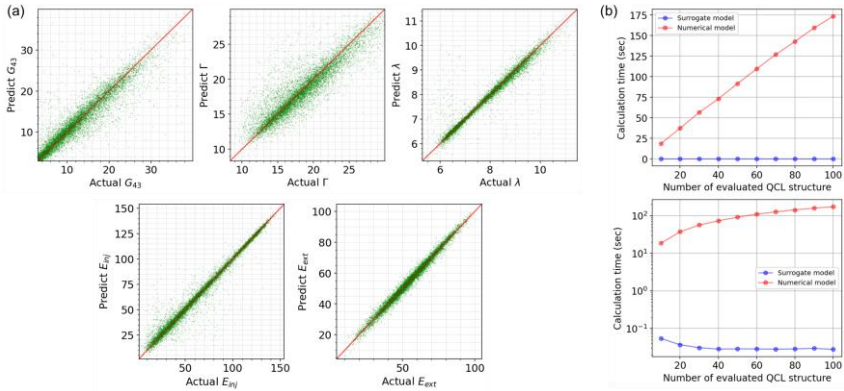

Figure 5 (a) Scatter plot between actual and predicted output characteristics of QCLs. (b) Optimization time comparison between DNN-TP surrogate model and numerical model in linear scale (upper), and logarithm scale (below).

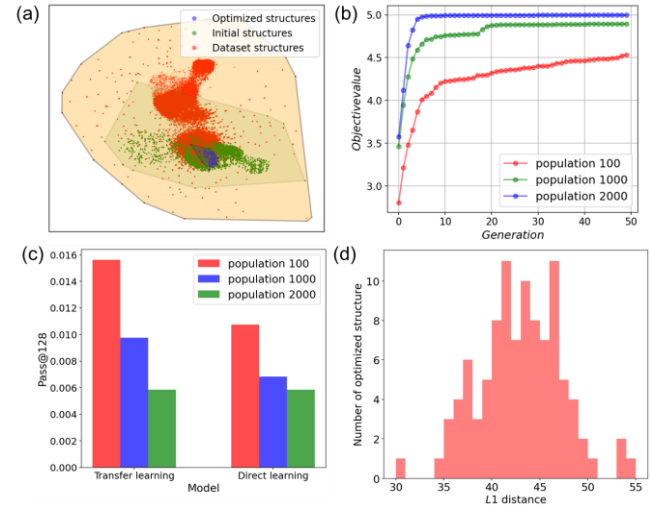

Figure 6 Optimization and evaluation results of the surrogate-assisted genetic algorithm for QCL design. (a) Convergence curve of the total objective score over generations for three different DNN training strategies: DL, GTL, and PBTL. (b) Pass@128 ratio comparison for each model, showing a significant improvement in optimization quality with PBTL. (c) Latent space visualization of the candidate structures throughout optimization, illustrating convergence from a dispersed to a compact distribution. (d) Histogram of L1 distances between the optimized and closest training dataset structures, indicating generation of novel designs beyond the training distribution.

While high exploration supports global search, it also increases the likelihood of out-of-distribution (OOD) queries, which can exacerbate the discrepancy between surrogate predictions and ground-truth physics. This observation underscores the importance of balancing exploration and surrogate model reliability in data-driven optimization. This efficiency enables extensive exploration of the design space while implementing a nature-inspired evaluation process into the photonic device optimization workflow—a key step toward biologically motivated optimization in quantum photonics.

### 3.3.3 Novelty and Physical Performance of Optimized structures

To investigate the novelty of the optimized designs, we calculate the L1 distance between each optimized structure and its closest counterpart in the training dataset. As shown in Figure 6(d), the average L1 distance for 100 verified optimized structures is approximately 43, with the minimum distance being 30. These results confirm that the surrogate-assisted GA is not merely reproducing training data but is capable of discovering genuinely novel QCL designs that extend beyond the training distribution.

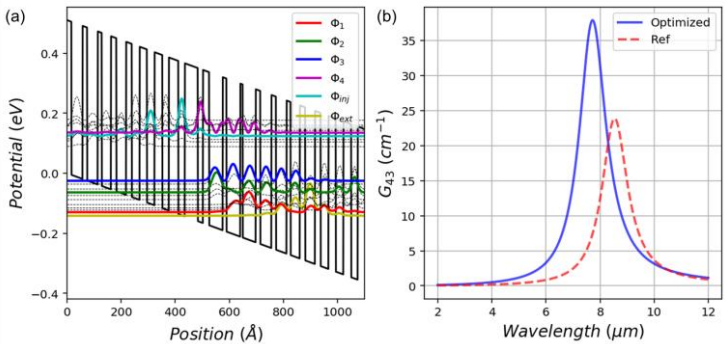

Figure 7 (a) Calculated energy levels and wavefunctions of a representative optimized QCL superlattice structure **48**/19/**23**/50/**14**/51/**9**/48/**17**/41/**17**/39/**15** /34/**20**/36/**21**/29/**27**/28/**29**/24/**35**/24 with thicknesses are in angstroms. The InAlAs barriers' thicknesses are bold, while the InGaAs wells are in normal face. (b) Modal gain spectra of the optimized structure and reference structure. The optimized structure showing a peak near the target wavelength.

Finally, the performance of one verified optimized structure is illustrated in Figure (a) and (b). The structure achieves a peak emission wavelength of 7.72 μm, a modal gain of 37.9 cm$^{-1}$, a linewidth of 12.5 meV, and effective injection and extraction energy levels of 11.42 meV and 4.98 meV, respectively. Compared to reference structure, the model gain is 54% increased and all the other parameters are within the operating range. These results confirm that the surrogate model, in conjunction with GA, can identify QCL designs that satisfy multi-objective criteria and maintain physical validity.

## 4. Conclusion

We proposed a NN-based surrogate modeling framework to efficiently optimize mid-IR QCL structures with five objectives. Given that the output latent space contains multiple local optima, a large population size per generation is required to ensure sufficient exploration. In this setting, the DNN-TP-assisted GA leverages a simple and parallelizable computation process, achieving over 80,000 times faster optimization compared to conventional numerical simulators. To train the DNN-TP model, we introduced a novel transfer learning strategy, PBTL, which incorporates domain-specific physical knowledge grounded in QCL operation principles. By training a core model (DNN-CP) in a reduced latent space, which contains compressed operation principle of the QCL emission process and transferring the learned representations to DNN-TP, we achieved a 0.69% improvement in prediction accuracy, corresponding to a 50% reduction in required training data (50,000 samples). While the numerical gain appears minor, it significantly affects optimization outcomes; notably, the pass@128 ratio of the PBTL model was 60% higher than that of the directly trained model.

Furthermore, we evaluated the L1 distance between physically verified optimized structures and their nearest neighbors in the training dataset, finding a minimum distance of approximately 30. This result demonstrates the surrogate model's capacity to generate novel and physically consistent designs beyond the training distribution, indicating a learned understanding of underlying quantum physics rather than simple memorization.

Although our method was demonstrated on QCL structures, it is broadly applicable to domains where feature importance is highly imbalanced, and data generation is expensive. The proposed framework—comprising a scalarized multi-objective GA, PBTL-based surrogate modeling, and biologically inspired optimization dynamics—offers a scalable and generalizable solution for real-time photonic design automation.

**Research funding:** This work was supported by the Basic Science Research Program (NRF-2021R1F1A1062591) through the National Research Foundation of Korea (NRF).

**Author contributions:** Gibaek Kim - Conceptualization, Methodology, Software, Visualization, Writing - original draft, Writing - review & editing. Jungho Kim - Conceptualization, Methodology, Visualization, Writing - original draft, Writing - review & editing.

**Conflict of interest:** The authors declare no conflict of interest.

**Data availability:** The datasets generated and analyzed during the current study are available from the corresponding author upon reasonable request.

## References

[1] J. Park, S. Kim, D. W. Nam, H. Chung, C. Y. Park, and M. S. Jang, "Free-form optimization of nanophotonic devices: from classical methods to deep learning," *Nanophotonics*, vol. 11, no. 9, pp. 1809–1845, 2022.

[2] W. Chen, S. Yang, Y. Yan, Y. Gao, J. Zhu, and Z. Dong, "Empowering nanophotonic applications via artificial intelligence: pathways, progress, and prospects," *Nanophotonics*, vol. 14, no. 4, pp. 429–447, 2025.

[3] S. So, T. Badloe, J. Noh, J. Bravo-Abad, and J. Rho, "Deep learning enabled inverse design in nanophotonics," *Nanophotonics*, vol. 9, no. 5, pp. 1041–1057, 2020.

[4] J. Faist, F. Capasso, D. L. Sivco, C. Sirtori, A. L. Hutchinson, and A. Y. Cho, "Quantum cascade laser," *Science*, vol. 264, no. 5158, pp. 553–556, 1994.

[5] C. Sirtori, F. Capasso, J. Faist, D. L. Sivco, A. L. Hutchinson, and A. Y. Cho, "Quantum cascade unipolar intersubband light emitting diodes in the 8–13 μm wavelength region," *Applied physics letters*, vol. 66, no. 1, pp. 4–6, 1995.

[6] R. Ferreira and G. Bastard, "Evaluation of some scattering times for electrons in unbiased and biased single- and multiple-quantum-well structures," *Phys. Rev. B*, vol. 40, no. 2, pp. 1074–1086, 1989.

[7] M. Hartig, S. Haacke, B. Deveaud, and L. Rota, "Femtosecond luminescence measurements of the intersubband scattering rate in Al$_x$Ga$_{1-x}$As/GaAs quantum wells under selective excitation," *Phys. Rev. B*, vol. 54, no. 20, pp. R14269–R14272, 1996.

[8] D. Mueller and G. Triplett, "Development of a multi-objective evolutionary algorithm for strain-enhanced quantum cascade lasers," *Photonics*, vol. 3, no. 3, p. 44, 2016.

[9] P. Ashok, "Particle swarm optimization approach to identify optimum electrical pulse characteristics for efficient gain switching in dual wavelength quantum cascade lasers," *Optik*, vol. 171, pp. 786–797, 2018.

[10] A. S. Dashkov and L. I. Goray, "QCL design engineering: automatization vs. classical approaches," *Semiconductors*, vol. 54, no. 14, pp. 1823–1825, 2020.

[11] M. Franckié and J. Faist, "Bayesian optimization of terahertz quantum cascade lasers," *Phys. Rev. Applied*, vol. 13, no. 3, p. 034025, 2020.

[12] Y. Qu, L. Jing, Y. Shen, M. Qiu, and M. Soljačić, "Migrating knowledge between physical scenarios based on artificial neural networks," *ACS Photonics*, vol. 6, no. 5, pp. 1168–1174, 2019.

[13] Z. Fan *et al.*, "Transfer-learning-assisted inverse metasurface design for 30% data savings," *Phys. Rev. Applied*, vol. 18, no. 2, p. 024022, 2022.

[14] G. Cho and J. Kim, "Effect of conduction band non-parabolicity on the optical gain of quantum cascade lasers based on the effective two-band finite difference method," *Semicond. Sci. Technol.*, vol. 32, no. 9, p. 095002, 2017.

[15] J. Kim *et al.*, "Theoretical and experimental study of optical gain and linewidth enhancement factor of type-I quantum-cascade lasers," *IEEE J. Quantum Electron.*, vol. 40, no. 12, pp. 1663–1674, 2004.
[16] T. Unuma, M. Yoshita, T. Noda, H. Sakaki, and H. Akiyama, "Intersubband absorption linewidth in GaAs quantum wells due to scattering by interface roughness, phonons, alloy disorder, and impurities," *J. Appl. Phys.*, vol. 93, no. 3, pp. 1586–1597, 2003.
[17] S. Tsujino *et al.*, "Interface-roughness-induced broadening of intersubband electroluminescence in p-SiGe and n-GaInAs/ AlInAs quantum-cascade structures," *Appl. Phys. Lett.*, vol. 86, no. 6, p. 062113, 2005.
[18] A. Wittmann, Y. Bonetti, J. Faist, E. Gini, and M. Giovannini, "Intersubband linewidths in quantum cascade laser designs," *Appl. Phys. Lett.*, vol. 93, no. 14, p. 141103, 2008.
[19] P. I. Frazier, "A tutorial on Bayesian optimization," Jul. 08, 2018, *arXiv*: arXiv:1807.02811. doi: 10.48550/arXiv.1807.02811.
[20] J. Park, S. Kim, D. W. Nam, H. Chung, C. Y. Park, and M. S. Jang, "Free-form optimization of nanophotonic devices: from classical methods to deep learning," *Nanophotonics*, vol. 11, no. 9, pp. 1809–1845, 2022.
[21] A. Bismuto, R. Terazzi, B. Hinkov, M. Beck, and J. Faist, "Fully automatized quantum cascade laser design by genetic optimization," *Appl. Phys. Lett.*, vol. 101, no. 2, p. 021103, 2012.
[22] A. Géron, *Hands-on machine learning with Scikit-Learn, Keras, and TensorFlow: Concepts, tools, and techniques to build intelligent systems*. O'Reilly Media, Inc., 2022.
[23] J. Kim *et al.*, "Inverse design of nanophotonic devices enabled by optimization algorithms and deep learning: recent achievements and future prospects," *Nanophotonics*, vol. 14, no. 2, pp. 121–151, 2025.
[24] C. Kang, C. Park, M. Lee, J. Kang, M. S. Jang, and H. Chung, "Large-scale photonic inverse design: computational challenges and breakthroughs," *Nanophotonics*, vol. 13, no. 20, pp. 3765–3792, 2024.
[25] A. Konak, D. W. Coit, and A. E. Smith, "Multi-objective optimization using genetic algorithms: A tutorial," *Reliab. Eng. Syst. Saf.*, vol. 91, no. 9, pp. 992–1007, 2006.
[26] T. Akiba, S. Sano, T. Yanase, T. Ohta, and M. Koyama, "Optuna: A next-generation hyperparameter optimization framework," ACM SIGKDD International Conference on Knowledge Discovery & Data Mining, 2019, pp. 2623–2631.
[27] R. Pestourie, Y. Mroueh, C. Rackauckas, P. Das, and S. G. Johnson, "Physics-enhanced deep surrogates for partial differential equations," *Nat. Mach. Intell.*, vol. 5, no. 12, pp. 1458–1465, 2023.
[28] W.-M. Lee, *Python machine learning*, New Jersey, John Wiley & Sons, 2019.
[29] P. Harrison and A. Valavanis, *Quantum wells, wires and dots: theoretical and computational physics of semiconductor nanostructures*, New Jersey, John Wiley & Sons, 2016.
[30] T. Hayes *et al.*, "Simulating 500 million years of evolution with a language model," *Science*, vol. 387, no. 6736, pp. 850–858, 2025.

# Supplementary Material

## 1 Numerical Modeling of QCLs

Quality of deep neural network (DNN) surrogate model is highly determined by a quality of training dataset. We use the effective two-band model at the zone center for conduction band non-parabolicity (NPB) [1]. In case of the effective two-band model in multi quantum well structures, wavefunctions and be expressed as (1), where $u_{c,v}(r)$ is Bloch function, $\exp(ik_t\rho)/\sqrt{A}$ is envelope function in the transverse direction, $\phi_{c,v}(z)$ indicates longitudinal envelop function. $k_t$ and $\rho$ represent transverse wavevector and position vector with respect to transverse dimension and represents area of the device.

$$\Psi(r) = u_c(r)\frac{\exp(ik_t\rho)}{\sqrt{A}}\phi_c(z) + u_v(r)\frac{\exp(ik_t\rho)}{\sqrt{A}}\phi_v(z) \quad (1)$$

The 2x2 Hamiltonian that can be implemented in the envelope function $\phi_{c,v}(z)$ is given in (2) at the zone center, where $E_{c,v}$ is position-dependent conduction/valence band edge, $m_0$ is effective mass of the electron, $p_{cv}$ is momentum matrix element, and $p_z$ is the momentum operator.

$$H = \begin{bmatrix} E_c(z) & \frac{p_{cv}}{m_0}p_z \\ -\frac{p_{cv}}{m_0}p_z & E_v(z) \end{bmatrix} \quad (2)$$

Thorough Eqs. (1) and (2) with $H\phi = E\phi$, we can obtain eigenvalues and eigenvectors that are corresponding to sub band energies and envelope functions. Using eigenvalues, we can obtain emission wavelength, injection energy and extraction energy through the equation below.

$$\lambda = 1.24/(E_4 - E_3) \quad (3)$$

$$\Delta E_{inj} = E_{inj} - E_4 \quad (4)$$

$$\Delta E_{ext} = E_1 - E_{ext} \quad (5)$$

Using those values, we can calculate the intersubband optical dipole moment, intersubband LO phonon scattering rate between state m and n (m>n) [2] thorough (3) and (4), respectively. $n_q = (\exp(\hbar\omega_{LO}/kt) - 1)^{-1}$ is a phonon occupation number, where k is Boltzmann constant, T is absolute temperature, and $\hbar\omega_{LO}$ is phonon energy. The term $\epsilon_p$ is effective dielectric constant of $\epsilon_p = 1/\epsilon_\infty - 1/\epsilon_s$, where $\epsilon_s$ is optical dielectric constant and $\epsilon_\infty$ is static dielectric constant.

$$\mu_{mn}^c = <\phi_c^n(z)|ez|\phi_c^m(z) > \hat{z} \quad (6)$$

$$\frac{1}{\tau_{mn}^{c(v)}} = \frac{m_e^* e^2 \omega_{LO}(n_q+1)}{4\hbar^2 \epsilon_p q_z} \times \int dz \int dz' \, \phi_{c(v)}^m(z)\phi_{c(v)}^n(z)e - qz|z - z'|\phi_{c(v)}^m(z')\phi_{c(v)}^n(z') \quad (7)$$

$$q_z = \sqrt{2m_e^*(E_m - E_n - \hbar\omega_{LO})/\hbar^2}$$

The intrasubband interface roughness (IFR) scattering rate is calculated by (8) [3,4,5], where $\Delta$ and $\Lambda$ are average roughness height and correlation length, respectively. $\delta U$ is the conduction band offset. Finally, we can obtain peak gain value and linewidth of QCL structures using (9) and (10). We can notice that it is necessary to employ both eigenvalues and eigenvectors in order to calculate scattering rates through above process.

$$\frac{1}{\tau_{IFR}} = \left(\frac{2\pi em^*}{\hbar^3}\right)\Delta^2\Lambda^2\delta U^2 \sum_k [\phi_m^2 - \phi_n^2]^2 \quad (8)$$

$$G_{43} = \frac{\Gamma\omega\pi}{en_r c\epsilon_0 L_p}|\mu_{43}|^2\tau_4\left(1 - \frac{\tau_3}{\tau_{43}}\right) \times \frac{J}{\pi\gamma_{43}} \quad (9)$$

$$\gamma = \frac{\hbar}{e}\left(\frac{1}{\tau_{POP}} + \frac{1}{\tau_{IFR}}\right) \quad (10)$$

## 2 Train dataset generation

QCL structures for each DNN predictors are randomly generated within the range displayed in Table S1 and S2. As each layer's scale differs from one another, min-max normalization is performed to ensure equal contribution. Also, there are some outliers in the dataset whose structure with maximum modal gain over 50 or optical linewidth of 120. We validated these data by analyzing their scattering time, eliminating physically unacceptable datasets.

Table S1. Data generation range of each layer for DNN-TP (unit: A).

|  | Start | End | # of candidate |
| --- | --- | --- | --- |
| Active 1 | 30 | 50 | 20 |
| Active 2 | 11 | 25 | 14 |
| Active 3 | 9 | 25 | 16 |
| Active 4 | 45 | 60 | 15 |
| Active 5 | 9 | 25 | 16 |
| Active 6 | 40 | 60 | 20 |
| Active 7 | 9 | 25 | 16 |
| Active 8 | 35 | 50 | 15 |
| Inject 1 | 16 | 30 | 15 |
| Inject 2 | 36 | 50 | 15 |
| Inject 3 | 11 | 20 | 10 |
| Inject 4 | 31 | 40 | 10 |
| Inject 5 | 11 | 20 | 10 |
| Inject 6 | 31 | 40 | 10 |
| Inject 7 | 11 | 20 | 10 |
| Inject 8 | 31 | 40 | 10 |
| Inject 9 | 14 | 23 | 10 |
| Inject 10 | 24 | 33 | 10 |
| Inject 11 | 24 | 33 | 10 |
| Inject 12 | 24 | 33 | 10 |
| Inject 13 | 26 | 40 | 15 |
| Inject 14 | 21 | 35 | 15 |
| Inject 15 | 31 | 45 | 15 |
| Inject 16 | 21 | 35 | 15 |

Table S2. Data generation range of each layer for DNN-CP (unit: A).

|  | Start | End | # of candidate |
| --- | --- | --- | --- |
| Active 1 | 30 | 50 | 20 |
| Active 2 | 11 | 25 | 14 |
| Active 3 | 9 | 25 | 16 |
| Active 4 | 45 | 60 | 15 |
| Active 5 | 9 | 25 | 16 |
| Active 6 | 40 | 60 | 20 |
| Active 7 | 9 | 25 | 16 |
| Active 8 | 35 | 50 | 15 |

Figure S1 illustrates the accuracy of the DNN-TP model increases as the number of training dataset size increases. However, we can see there are obvious saturation around 100,000 dataset size. Considering computational efficiency, we set training dataset size for DNN-TP as 100,000.

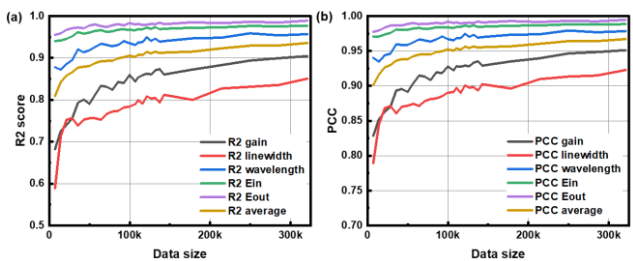

Figure S1. R2 score and PCC of DNN-TP with varying training dataset size. DNN-TP is trained with direct learning methodology during this examination.

## 3 Training of DNN-CP and DNN-TP with direct learning methodology.

In this section, we present the training details of DNN-CP and DNN-TP using the direct learning methodology. In the case of DNN-CP, it consists of 500-500-500-500-250 nodes and each node are connected with Leaky Relu activation function. For the DNN-TP, we use NN structure with 500-500-500-500-250-125 nodes given high complexity of its task. To mitigate overfitting, we use early stopping method with patience of 50. The other hyperparameters are identical as presented in Table S3.

Table S3. Training Hyperparameters of two model.

|  | DNN-CP | DNN-TP |
| --- | --- | --- |
| Optimizer | Adam | Adam |
| Learning rate | 1e-4 | 1e-4 |
| Batch size | 32 | 32 |
| Initialization | Xavier | Xavier |

The training results of the two neural networks differ significantly due to their different input feature dimensionalities and dataset sizes (8 and 24, respectively). As mentioned in the main text, DNN-CP exhibits better training performance than DNN-TP, likely because of its lower input dimensionality. Similarly, Figure S2(a) exhibits a clearer linear trend than Figure S2(b). We also examined the PCC of each model trained with the direct learning method. Table S4 shows that DNN-CP is more accurate than DNN-TP. This implies that DNN-CP's high representation ability, learned from a narrow feature domain, can improve the prediction accuracy of DNN-TP through transfer learning.

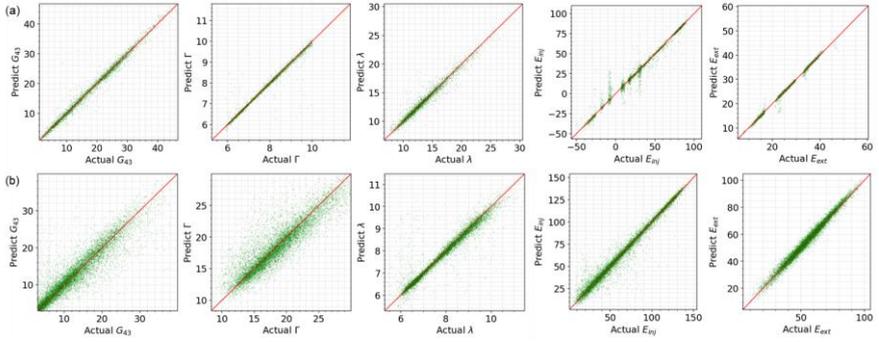

Figure S2. Comparison of predicted versus actual values using scatter plots for (a) DNN-CP and (b) DNN-TP models, trained with direct learning methodology.

Table S3. PCC of prediction of trained models.

|  | DNN-CP | DNN-TP |
| --- | --- | --- |
| Modal Gain | 0.989 | 0.913 |
| Linewidth | 0.988 | 0.890 |
| Emission wavelength | 0.985 | 0.966 |
| Effective injection energy | 0.985 | 0.984 |
| Effecitve extraction energy | 0.997 | 0.990 |

## 4 Optimization of transfer learning methodology

The effect of the transfer learning depends on the combination of the transferred layers. To optimize transfer learning methods, the R2 score of each transfer combination is examined. The results of this optimization process are presented in Figure S3. Due to discrepancies between the two neural networks (8 input dimensions for DNN-CP and 24 input dimensions for DNN-TP), GTL can only use hidden layers 2 and 3 as transferred layers. In contrast, PBTL has the exact same input dimensions as the DNN-CP, allowing the transfer of a wider range of information. The findings of the optimization indicate that transferring the third layer of the hidden layer is the optimal option in both methods. In most cases, the GTL results in degraded accuracy, while the PBTL yields increased accuracy. The degradation of GTL can be attributed to the distortion of input information from an unsuitable transfer method. Therefore, we can say that it is important to have sufficient domain knowledge when using transfer learning to avoid reducing model performance.

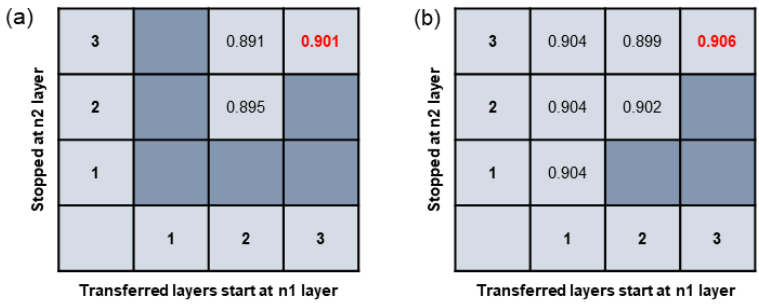

Figure S3. Optimization result for two transfer learning methodolgies: PBTL, and GTL.